\documentclass[aps,prc,twocolumn,showpacs,showkeys,preprintnumbers,amsmath,amssymb,a4paper]{revtex4-1}

\usepackage{amssymb}
\usepackage{amsmath, bbm}
\usepackage[figuresright]{rotating}
\usepackage{capt-of}

\newcommand{\La}{{\Lambda}}
\newcommand{\Si}{{\Sigma}}

\newcommand{\be}{\begin{eqnarray}}
\newcommand{\ee}{\end{eqnarray}}

\newlength{\feynwidth} 
\newlength{\feynwidthbig} 

\usepackage{xcolor} 

\begin{document}

\title{On the structure in the $\Lambda N$ cross section at the $\Sigma N$ threshold}

\author{Johann Haidenbauer$^{1}$, Ulf-G. Mei{\ss}ner$^{2,1,3}$}

\affiliation{
$^{1}${Institute for Advanced Simulation, Institut f\"ur Kernphysik and J\"ulich Center for Hadron Physics,
  Forschungszentrum J{\"u}lich, D-52425 J{\"u}lich, Germany}
\\
$^{2}${Helmholtz-Institut f\"ur Strahlen- und Kernphysik and Bethe Center for Theoretical Physics,
  Universit\"at Bonn, 53115 Bonn, Germany}
\\
$^{3}${Tbilisi State  University,  0186 Tbilisi, Georgia}
}
\begin{abstract}
The complexity of threshold phenomena is exemplified on a prominent and long-known case - 
the structure in the $\Lambda p$ cross section (invariant mass spectrum) at the opening of the 
$\Sigma N$ channel.
The mass splitting between the $\Sigma$ baryons together with the angular momentum coupling in the
$^3S_1$-$^3D_1$ partial wave imply that, in principle, up to six channels are involved. 
Utilizing hyperon-nucleon potentials that provide an excellent description of the available low-energy
$\Lambda p$ and $\Sigma N$ scattering data, the shape of the resulting $\Lambda p$ cross section is 
discussed and the poles near the $\Sigma N$ threshold are determined. Evidence for 
a strangeness $S=-1$ dibaryon is provided, in the form of a deuteron-like 
(unstable) $\Sigma N$ bound state. 
Predictions for level shifts and widths of $\Sigma^-p$ atomic states are given. 
\end{abstract}

\maketitle

\section{Introduction} 

There has been a revival of interest in threshold phenomena over the last two 
decades triggered not least by the discovery of the so-called XYZ states 
\cite{Guo:2018,Brambilla:2020}.  
Many of the new states seen in the charm or bottom sectors, which do not fit into the
standard quark-antiquark or three-quark classification, have been observed close 
to thresholds. The perhaps most famous one is the $\chi_{c1} (3872)$, formerly known 
as $X(3872)$, whose mass coincides with the $D \bar D^*$/$ D^* \bar D$ threshold 
within the experimental uncertainty \cite{PDG}. Threshold anomalies are standard 
textbook knowledge for a long time, see e.g. Ref.~\cite{Newton:1966}, but have 
been revisited in various works since they are considered as possible 
explanation for the nature and structure of some XYZ and other 
exotic states, see, e.g. \cite{Matuschek:2021,Dong:2021} and the 
reviews \cite{Yamaguchi:2020,Guo:2020} for very recent examples.

In the light of this development and also because of its intrinsic interest we re-examine
one of the longest known threshold effects, namely the structure observed in 
the $\La p$ cross section (invariant mass spectrum) at the threshold of the $\Si N$ 
channel. First evidence of it was already reported as early as 1961 \cite{Dahl:1961}, 
but the first convincing signal and still one of the most prominent examples 
is from the measurement of the reaction $K^- d \to \pi^- \La p$ by 
Tan in 1969 \cite{Tan:1969}. A review of other early observations can be found in 
Ref.~\cite{Dalitz:1980} and an overview of later measurements
is provided by Machner et al.~\cite{Machner:2013}. 
More recent examples for the presence of a $\Si N$ threshold effect, in the reaction 
$pp \to K^+ \La p$, can be found in 
Refs.~\cite{ElSamad:2013,Roder:2013,Hauenstein:2017,Munzer:2018}. 
Very recently evidence of the threshold structure has been also observed 
in measurements of 
the $\La p$ correlation function in $pp$ collisions at $T=13$~GeV by the 
ALICE Collaboration \cite{Acharya:2021}. 
Let us mention that there are also data for the $\La p$ elastic cross section itself 
in the $\Si N$ threshold region \cite{Kad71,Hau77}. However, in that case the energy 
resolution is rather poor so that no conclusion on a possible structure could be 
drawn \cite{Machner:2013}. 

In the present work we discuss the predictions for $\La N$ observables around the 
$\Si N$ threshold, utilizing hyperon-nucleon ($YN$) interactions that yield the presently 
best description of low-energy $\La p$ \cite{Sec68,Ale68}, 
$\Si^- p$ \cite{Eng66,Eis71,Hep68,Ste70} and $\Si^+ p$ \cite{Eis71} scattering data. 
This condition is met by the $YN$ potentials derived within chiral effective 
field theory (EFT) up to next-to-leading order (NLO) by the J\"ulich-Bonn-Munich 
group \cite{Haidenbauer:2013,Haidenbauer:2019} and by the 
Nijmegen NSC97 meson-exchange potentials \cite{Rijken:1999}. In all those cases
the achieved $\chi^2$ value is in the order of $16$ for the $36$ (or $35$) ``best''
$YN$ data taken into account. 
The coupling of the $\Si N$ channel to $\La N$ dominates the dynamics
around the $\Si N$ threshold, where the angular-momentum coupled 
partial waves $^3S_1$ and $^3D_1$ play an important role. A further facet is 
added by the mass splitting between the $\Sigma$ baryons which implies that 
there are actually two physical $\Si N$ thresholds so that, in principle,  
one faces a six-channel problem. Thus, the $\La N$-$\Si N$
system is an excellent textbook example to illustrate issues and complications 
of coupled-channel dynamics that might be also instructive for interpreting 
threshold structures seen in the charm and/or bottom sector. 

We also re-address the question regarding a strangeness $S=-1$, isospin-$1/2$,
spin-$1$ dibaryon~\cite{Oakes:1963,Jaffe:1977}, building on those aforementioned 
$YN$ interactions. Indeed, the dispute about whether there is a dibaryon - in form 
of a deuteron-like $\Si N$ bound state - or not has a varied history. 
In the past, studies where the possibility of such a dibaryon 
was discussed focused primarily on the reaction
$K^- d \to \pi^- \La p$ 
\cite{Dosch:1978,Dosch:1980,Toker:1979,Toker:1981,Dalitz:1980,Dalitz:1983,Torres:1986,Deloff:1989}.
While initial investigations were more or less inconclusive, in the latest 
works \cite{Toker:1981,Torres:1986,Deloff:1989} the unanimous conclusion has been drawn 
that a $\Si N$ bound state does not exist near the $\Si N$ threshold. 
However, one must keep in mind that in those studies simplified models of 
the $YN$ interaction were employed. Specifically, 
with regard to the $^3S_1$-$^3D_1$ partial wave where that dibaryon
should occur, the tensor coupling mediated by the long-ranged one-pion exchange 
was ignored and usually only the $S$-wave component was taken into account. 
Realistic $YN$ potentials suggest that the $\La N \to \Si N$ transition occurs 
predominantly from the $\La N$ $^3D_1$ state  
\cite{Haidenbauer:2013,Haidenbauer:2019,Rijken:1999,Polinder:2006,Nagels:1977,Nagels:1979,Holzenkamp:1989,Reuber:1993,Haidenbauer:2005}.
Furthermore, often no constraints from SU(3) flavor symmetry were implemented.  
Though SU(3) symmetry is certainly broken, may be on the level of 
20 - 30\,\% \cite{Dover:1990}, one should not abandon it altogether. 

The paper is structured in the following way: 
In Sect.~II we provide the main results of our study. 
We start with a brief description of the employed $YN$ potentials 
and summarize the achieved $\chi^2$.  
Then we examine in detail the $\La p$ cross section near 
the $\Si N$ threshold based on three selected $YN$ interactions, 
the chiral EFT potentials NLO13~(600) \cite{Haidenbauer:2013} and 
NLO19~(600) \cite{Haidenbauer:2019}, and the 
Nijmegen NSC97f potential \cite{Rijken:1999}, 
in order to expose subtle differences in the dynamics.
Subsequently, we determine the pole positions in the complex plane 
near the $\Si N$ thresholds for the $^3S_1$-$^3D_1$ partial wave, 
and we discuss possible evidence for a dibaryon.
Finally, we compare the structures in the $\La p$ and $\La n$ 
cross sections around the $\Si N$ threshold. 
In Sect.~III predictions for the level shifts and widths of $\Si^-p$ atomic
states are given. 
The paper ends with a brief summary.

\section{Results} 

\subsection{The $YN$ potentials} 
Let us start by noting that $YN$ potentials which include the 
$\La N$-$\Si N$ coupling 
\cite{Haidenbauer:2013,Haidenbauer:2019,Rijken:1999,Polinder:2006,Nagels:1977,Nagels:1979,Holzenkamp:1989,Reuber:1993,Haidenbauer:2005}
are established by considering data from channels with different charge $Q$, namely 
those for $Q=0$ ($\Si^-p$), $Q=1$ ($\La p$), and $Q=2$ ($\Si^+p$). 
Thus, the predictions of the potentials
for $\La p$ near the $\Si N$ threshold, and specifically of the threshold
structure, are actually predominantly determined by the available cross sections
for the $\Si^-p$ elastic \cite{Eis71} and charge exchange ($\Si^-p\to \Si^0 n$) \cite{Eng66}
channels, and the one of the transition $\Si^-p\to\La n$ \cite{Eng66}. 
In addition there is the capture ratio at rest \cite{Hep68,Ste70}. 
The latter is defined by \cite{deSwart:1962}
\begin{eqnarray}
\nonumber
   r_R
      &=&\frac{1}{4}\,\frac{\sigma_s(\Sigma^-p\rightarrow\Sigma^0n)}
                    {\sigma_s(\Sigma^-p\rightarrow\Lambda n)
                    +\sigma_s(\Sigma^-p\rightarrow\Sigma^0n)} \\
      &+&\frac{3}{4}\,\frac{\sigma_t(\Sigma^-p\rightarrow\Sigma^0n)}
                    {\sigma_t(\Sigma^-p\rightarrow\Lambda n)
                    +\sigma_t(\Sigma^-p\rightarrow\Sigma^0n)}\,,
\label{rR}
\end{eqnarray}
where $\sigma_s$ ($\sigma_t$) is the total reaction cross section in the singlet
$^1S_0$ (triplet $^3S_1$-$^3D_1$) partial wave. The cross sections are the ones 
at zero momentum, but in calculations it is common practice \cite{Rijken:1999} 
to evaluate the cross sections at a small non-zero momentum,
namely $p_{\rm lab}=10$~MeV/c.
As mentioned already,
available $\La p$ cross sections in the $\Si N$ threshold region \cite{Kad71,Hau77}
are afflicted by a poor momentum resolution and usually not taken into account
in the fitting procedure.
 
For a detailed description of the utilized $YN$ interactions (NLO13, NLO19, NSC97f) 
we refer the reader to the original publications 
\cite{Haidenbauer:2013,Haidenbauer:2019,Rijken:1999}. 
Here we focus only on the essential features and differences. For all potentials 
SU(3) flavor symmetry is used as an essential guideline in the derivation. However, 
in the actual calculations it is broken in various ways, notably by the mass
differences of the pseudoscalar mesons $\pi$, $\eta$ and $K$. 
In the NSC97 potentials there is also an explicit SU(3) breaking in the
baryon-baryon-meson coupling constants. 
In the chiral EFT potentials there is no additional breaking of SU(3) symmetry. 
In particular, the short-distance dynamics, represented in that approach by 
contact terms, fulfills strict SU(3) symmetry in the original potential 
(NLO13) \cite{Haidenbauer:2013} and also in the version from 2019 
(NLO19) \cite{Haidenbauer:2019}. 
However, in both cases there is an explicit SU(3) symmetry breaking with
respect to the $NN$ system. 
 
In the EFT interactions the empirical binding energy of the hypertriton $^3_\La \rm H$
is used as a further constraint. It is utilized to fix the relative strength of the
spin-singlet and spin-triplet $S$-wave contributions to the $\Lambda p$ interaction.
Because of that, all NLO interactions yield practically identical values for
the $^1S_0$ and $^3S_1$ scattering length, respectively. In the NSC97 potentials 
there is no such constraint and, consequently, there is a fairly large variation in 
the $\Lambda p$ scattering lengths for the versions a-f presented in
Ref.~\cite{Rijken:1999}, correlated with the magnitude of explicit SU(3) symmetry
breaking. 
Anyway, as we will see below, this aspect has very little influence on
the $\Lambda p$ results near the $\Si N$ threshold.
Finally, we want to mention that isospin symmetry is fulfilled by the
EFT potentials \cite{Haidenbauer:2013,Haidenbauer:2019}. 
In case of the NSC97 interactions there is an isospin breaking in the
$\La N$ sector via $\Si^0$-$\La$ mixing \cite{Dalitz:1964} which 
allows for contributions from the exchange of isovector mesons ($\pi$, $\rho$, ...) 
to the $\La N \to \La N$ potential. Also this is relevant only for the $\La N$
results at low energies but not at the $\Si N$ threshold.
 
Note that a regularization is required when solving the scattering equation for 
interactions derived from chiral EFT \cite{Epelbaum:2009} which is usually 
done by introducing an exponential regulator function involving a cutoff. 
In case of the EFT potentials employed in the present study cutoff values 
of $500 - 650$~MeV have been used \cite{Haidenbauer:2013,Haidenbauer:2019} 
and we present here results for that range. As we will see, there is a  
small but noticeable cutoff dependence.

\begin{table*}
\caption{Achieved $\chi^2$ for the NLO13 \cite{Haidenbauer:2013} and 
NLO19 \cite{Haidenbauer:2019} interactions (for cutoffs $500$-$650$~MeV) 
and the J\"ulich '04 \cite{Haidenbauer:2005} and Nijmegen NSC97f \cite{Rijken:1999}
$YN$ potentials. The set includes 24 $\Si N$ data points.
For J\"ulich '04 the result without $r_R$ is given in brackets.
}
\renewcommand{\arraystretch}{1.2}
\label{tab:chi1}
\vspace{0.2cm}
\centering
\begin{tabular}{|c|rrrr|rrrr|r|r|}
\hline
reaction & \multicolumn{4}{|c|}{NLO13} & \multicolumn{4}{|c|}{NLO19} & J\"ulich '04 & NSC97f  \\
\hline
                & $500$ & $550$ & $600$ & $650$ & 
 $500$ & $550$& $600$& $650$  & & \\
\hline
\hline
$\Sigma^- p \to \La n$ \cite{Eng66} &$3.7$  &$3.9$ &$4.1$ &$4.4$ 
 &$4.7$  &$4.7$ &$4.0$ &$4.4$  &$8.3$ &$3.9$\\
$\Sigma^- p \to \Sigma^0 n$ \cite{Eng66} &$6.1$  &$5.8$ &$5.8$ &$5.7$ 
 &$5.5$  &$5.5$ &$6.0$ &$5.7$  &$6.4$ &$6.0$\\
$\Sigma^- p \to \Sigma^- p$ \cite{Eis71} &$2.0$  &$1.8$ &$1.9$ &$1.9$ 
 &$3.0$  &$2.9$ &$2.2$ &$1.9$  &$1.6$ &$2.3$\\
$\Sigma^+ p \to \Sigma^+ p$ \cite{Eis71} &$0.3$  &$0.4$ &$0.5$ &$0.3$ 
 &$0.3$  &$0.4$ &$0.4$ &$0.3$  &$0.1$ &$0.2$\\
\hline
$r_R$ \cite{Hep68,Ste70} &$0.1$  &$0.2$ &$0.1$ &$0.2$ 
 &$1.1$  &$0.7$ &$0.1$ &$0.5$  &$53.6$ &$0.0$\\
\hline
 total $\chi^2$                &$12.2$ &$12.0$ &$12.3$ &$12.5$ 
 &$14.6$  &$14.2$ &$12.7$ &$12.8$  &$70\,(16.4)$ &$12.4$\\
\hline
\hline
\end{tabular}
\renewcommand{\arraystretch}{1.0}
\end{table*}

Because of the important role played by the $\Si N$ data we summarize
the relevant $\chi^2$ values for the potentials considered in the present 
work in Table~\ref{tab:chi1}. The best results achieved correspond to 
$\chi^2 \approx 12-13$ for the $24$ $\Si N$ data points included. The 
NLO19 interactions with cutoffs of $500$ and $550$~MeV deviate already 
noticeably from the best values, which has consequences as we will see later. 
In case of the J\"ulich~'04 interaction, considered here for illustration,
the $\chi^2$ is very large, though mostly due to the fact that the capture
ratio $r_R$, which has been determined to very high precision \cite{Hep68,Ste70}, 
was not included in the fitting procedure. 
The Nijmegen potentials NSC97a-e yield a $\chi^2$ very close to 
that of NSC97f \cite{Rijken:1999}.  

\subsection{$\La p$ cross sections}

\begin{figure*}
\begin{center}
\includegraphics[height=67mm]{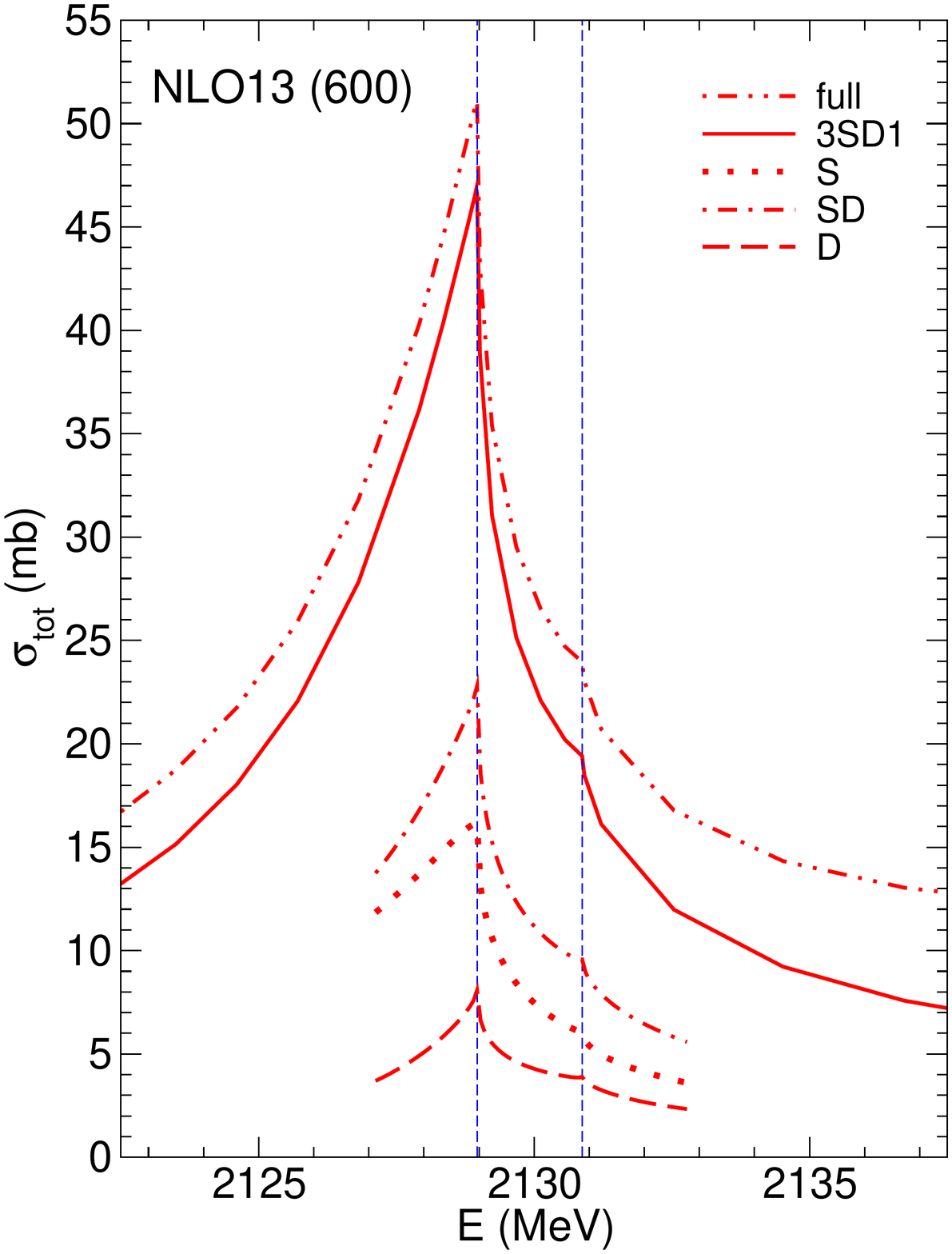}\includegraphics[height=67mm]{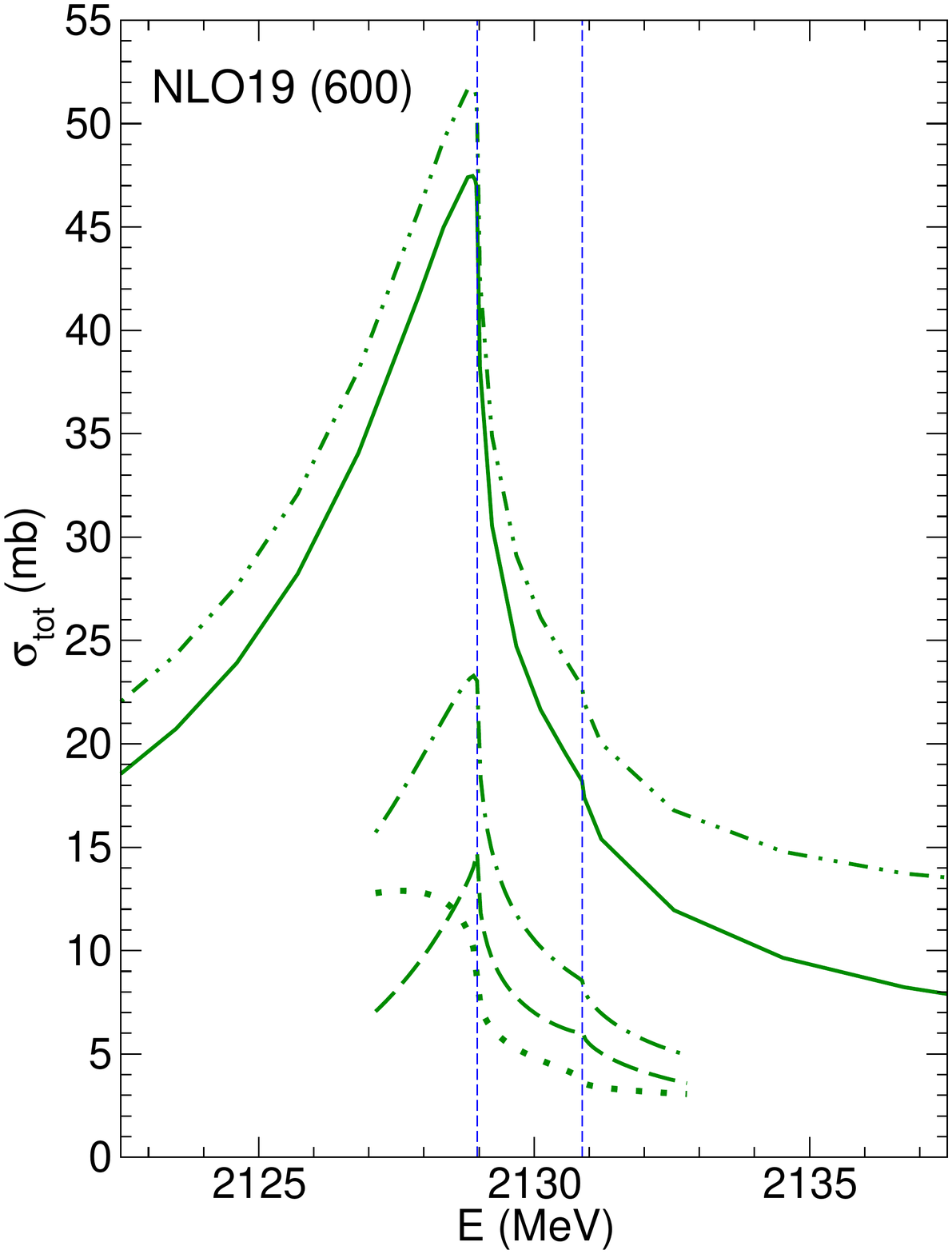}\includegraphics[height=67mm]{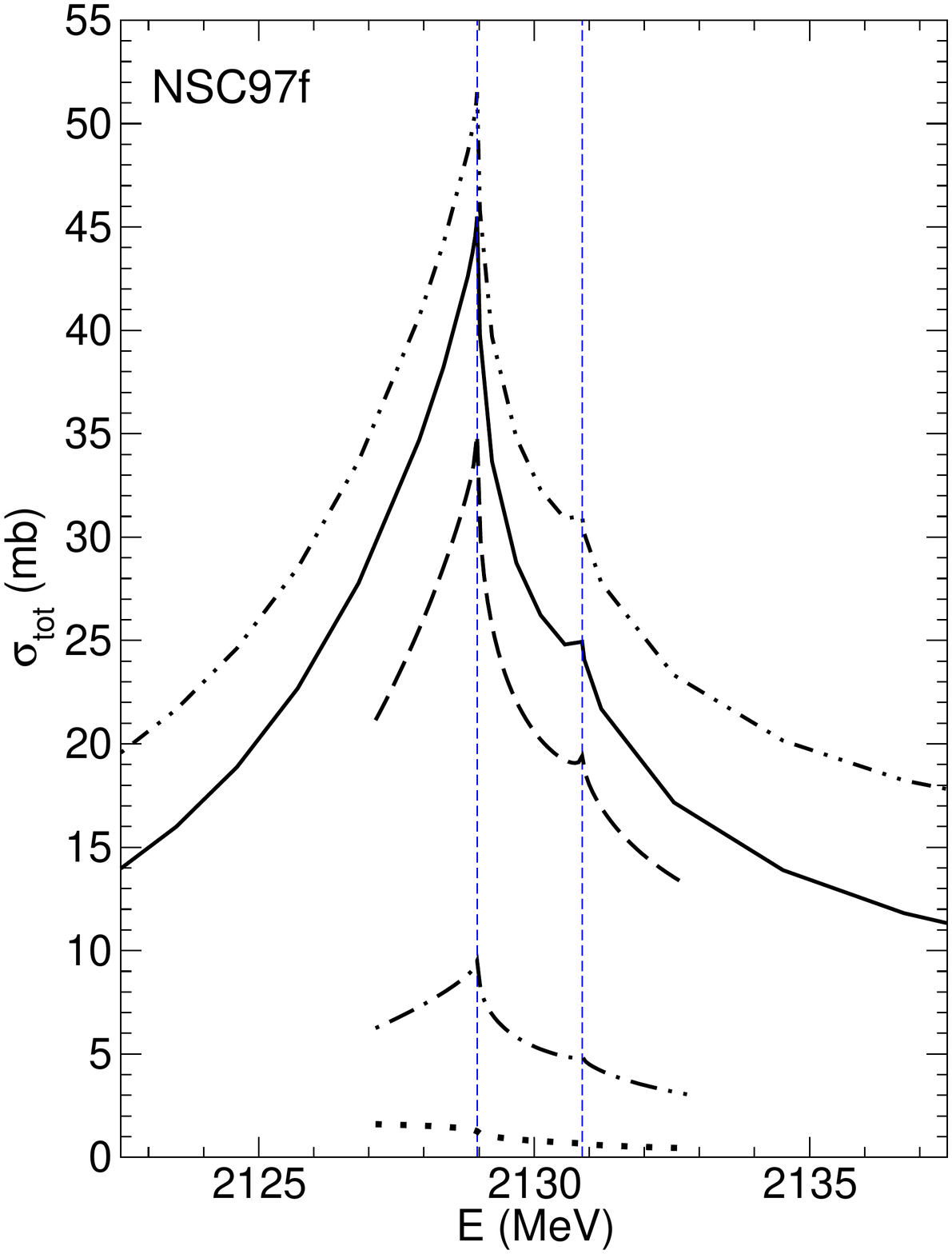}
\caption{Partial-wave contributions to the $\La p$ cross section for NLO13~(600) (left), 
NLO19~(600) (middle) and Nijmegen NSC97f (right) around the $\Si N$ threshold. 
The dash-double-dotted line is the full result, while the solid line is that of
the $^3S_1$-$^3D_1$ partial wave alone. 
The dotted, dash-dotted, and dashed lines represent the individual contributions from 
the $^3S_1$, the $S-D$ transition, and from the $^3D_1$ amplitudes, respectively.
The vertical lines indicate the $\Si^+ n$ and $\Si^0 p$ thresholds, respectively. 
}
\label{fig:Cusp}
\end{center}
\end{figure*}

It is instructive to first look  at the $\La p$ cross section in the $\Si N$ threshold 
region. Corresponding predictions are presented in 
Fig.~\ref{fig:Cusp} for the NLO13~(600), the NLO19~(600), and the NSC97f 
potentials. The dash-double-dotted lines representing the full results make clear 
that the cross sections are remarkably similar, especially close to the threshold of 
the lower channel ($\Si^+ n$). 
Furthermore, for all three potentials the $^3S_1$-$^3D_1$ partial wave 
(cf.  the solid lines) is responsible for about 90\% of the cross section.
Differences in the dynamics are reflected primarily in the individual components of this 
angular-momentum coupled partial wave, i.e. the $^3S_1$ (dotted), the $^3D_1$ (dashed) 
and the $S$-$D$ (dash-dotted) transition amplitude. 
Obviously, the NSC97f result is by far dominated by the $^3D_1$ amplitude while for 
the EFT interactions the largest contribution comes from the $S$-$D$ transition amplitude. 
Moreover,
there are differences in the very details. In case of NLO13 all three contributions
exhibit a cusp-like structure at the $\Si^+ n$ threshold. On the other hand,
for NLO19 and NSC97f a rounded step \cite{Badalyan:1982} appears in the $^3S_1$ amplitude, 
which is clearly visible for the EFT interaction but difficult to see for NSC97f 
because for the latter the contribution of the $^3S_1$ is fairly small. 

\begin{figure*}
\begin{center}
\includegraphics[height=67mm]{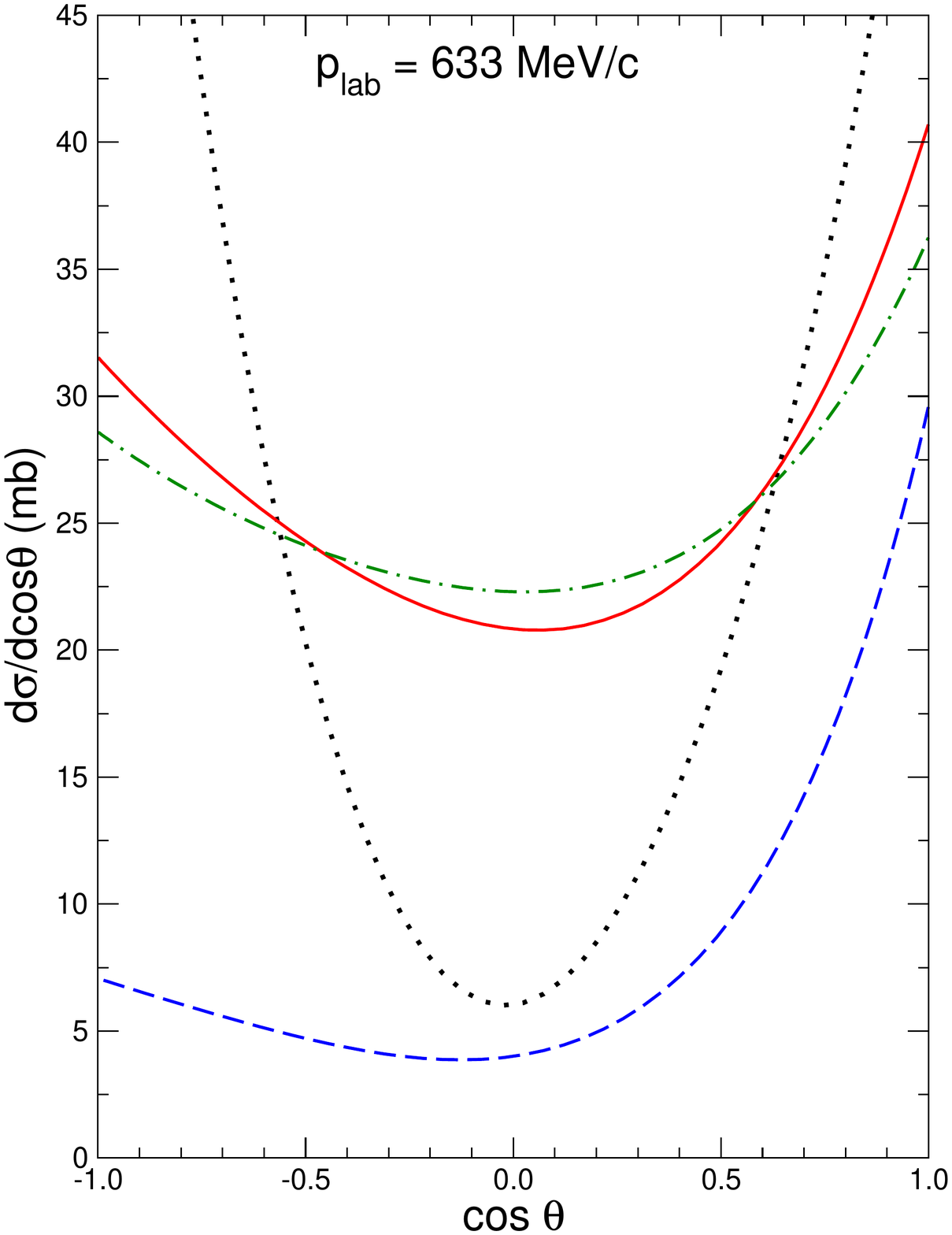}\includegraphics[height=67mm]{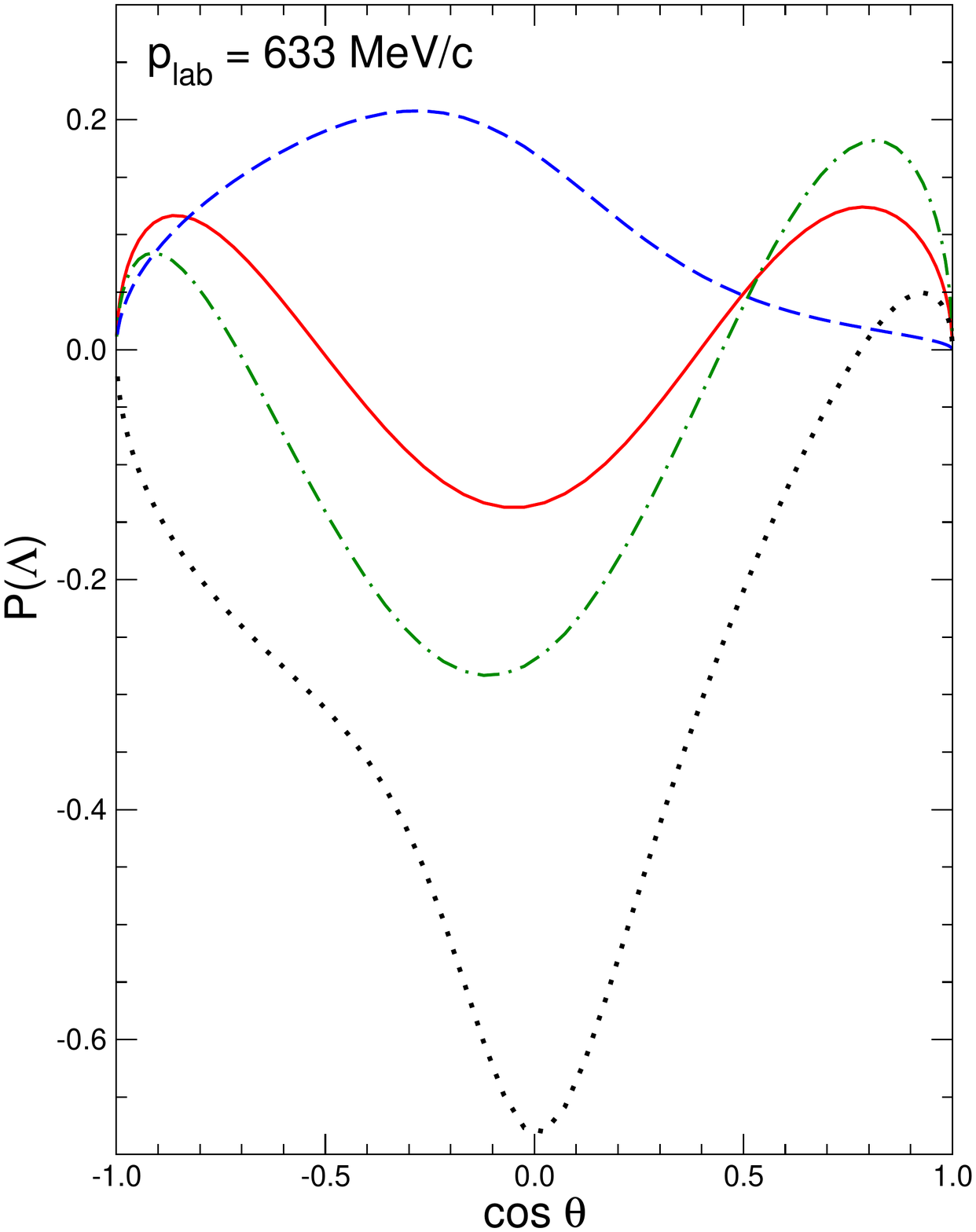}\includegraphics[height=67mm]{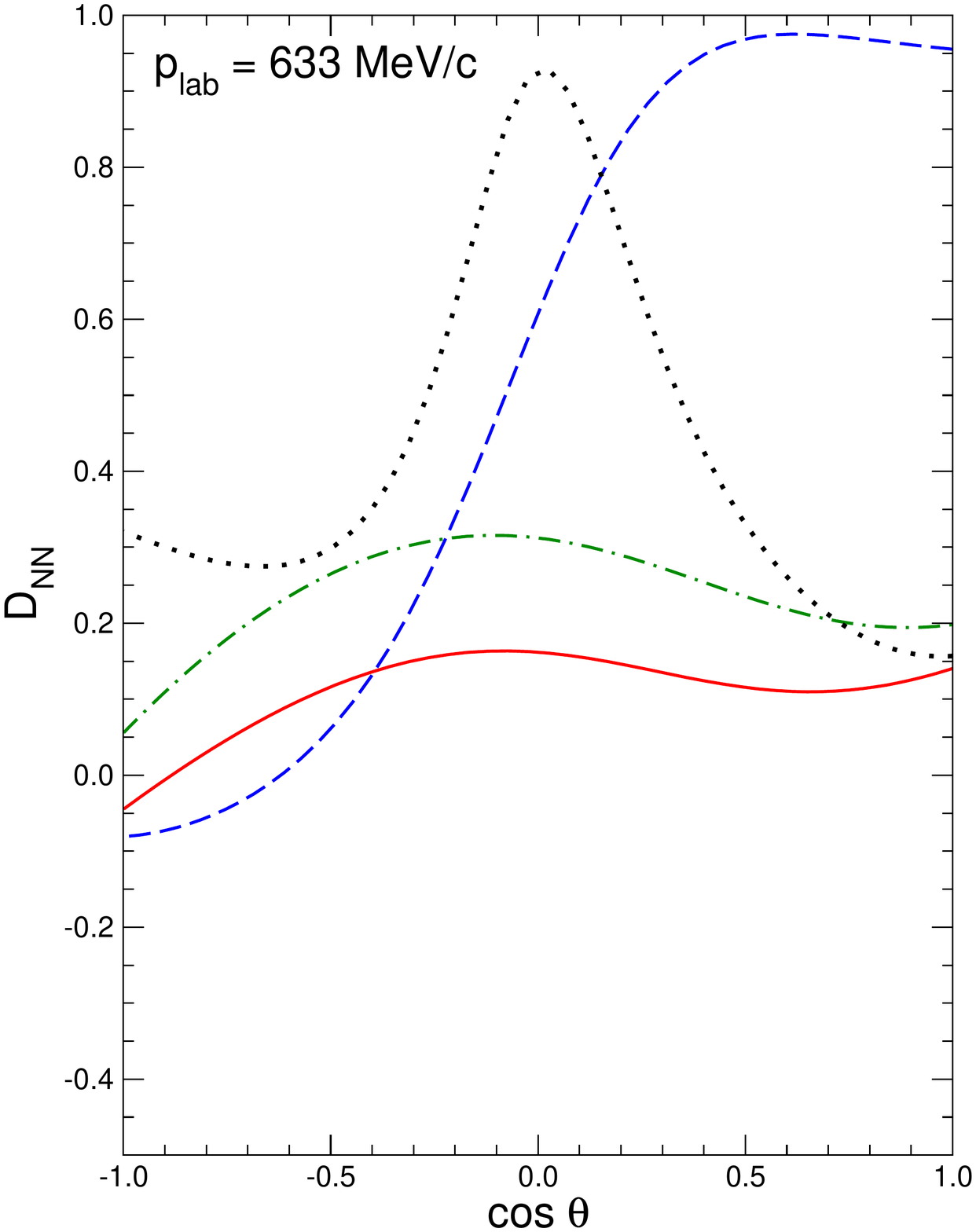}
\caption{Differential cross sections, polarization, and depolarization $D_{NN}$ for $\La p$
scattering at $633$~MeV/c (i.e. at the $\Si^+ n$ threshold). 
Predictions for NLO13~(600) (solid line), NLO19~(600) (dash-dotted), J\"ulich '04 (dashed),
and Nijmegen NSC97f (dotted) are presented. 
}
\label{fig:DN4}
\end{center}
\end{figure*}

Evidently, with a measurement of the $\La p$ cross section across the $\Si N$ 
threshold region, even with an excellent energy resolution, it will be difficult 
to resolve the dynamical differences represented by these scenarios. 
The only promising tool for a discrimination are measurements of differential observables.
This is exemplified in Fig.~\ref{fig:DN4} with predictions at $p_{lab}=633$~MeV/c, 
i.e. at the $\Si^+ n$ threshold. Of course, the observables are
also influenced by the $P$ waves (and higher partial waves) which
have uncertainties, too \cite{Polinder:2006,Haidenbauer:2013}. 
However, as one can see, there is definitely a qualitative difference 
in case where the $^3D_1$ contribution is dominant. Note that the NLO13 and NLO19 potentials
differ only in the $^1S_0$ and $^3S_1$-$^3D_1$ partial waves \cite{Haidenbauer:2019}. 
The interactions in the higher partial waves are identical. 
With regard to the J\"ulich '04 potential it should be said that its $\La p$ cross section 
in the $\Si N$ threshold region differs drastically from those shown in Fig.~\ref{fig:Cusp}  
\cite{Haidenbauer:2005}. 
Let us mention that measurements of differential observables for $\La p$ are planned at 
J-PARC \cite{JPARC:2020,Miwa}, also for energies in the $\Si N$ threshold region. 

In this context we want to emphasize that the subtle differences discussed above will have 
an impact on the outcome for reactions like $K^-d \to \pi^- \La p$ and/or $pp \to K^+\La p$, too.
Depending on the (principally unknown) reaction mechanism the relative weight of the $S$ 
and $D$ waves will differ in those processes as compared to $\La p$ elastic 
scattering. Accordingly, the structure or line shape in elastic scattering and in the 
$\La p$ invariant mass spectrum can certainly be different. 
Most of the past studies of $K^-d \to \pi^- \La p$ relied on $\La p$ interactions
that include only the $^3S_1$ partial wave \cite{Dosch:1978,Dosch:1980,Toker:1979,Toker:1981} 
and, thus, the above aspect is not accounted for. 

\subsection{Pole positions}

Let us now come to the pole positions for the $\La N$-$\Si N$ system in the 
$^3S_1$-$^3D_1$ partial wave. We determine those based on the multichannel 
effective range expansion discussed, e.g., in
Refs.~\cite{deSwart:1962,Shaw:1962,Badalyan:1982}
and also in the textbook by Newton~\cite{Newton:1966}. Such methods are
also used in lattice QCD to determine resonance parameters, see e.g.
\cite{Agadjanov:2016fbd}.
In the single-channel case the effective range expansion (ERE) of the 
scattering amplitude $f(q)= (S-1) / 2 i q$ is introduced via 
$f(q) = 1/(q\cot\delta - iq)$, where $q\cot\delta= -1/a + r\,q^2 / 2 + ...$.
Here, $S$ is the $S$-matrix, $q$ is the on-shell momentum, $\delta$ is the
phase shift and $a$ and $r$ are the scattering length and the effective range,
respectively. 
In the multichannel case the $S$-matrix is connected with the scattering matrix
${F}$ via $S_{ij} = \delta_{ij} + 2i \sqrt{q_iq_j} F_{ij}$, where the indices
$i$ and $j$ denote the channels and coupled partial waves. 
The scattering amplitude $F$ can be written in matrix form as
\begin{equation}
F = [{M} - {\rm i}\, q]^{-1}, \quad M = -A^{-1} + R\, q_0^2/2 + P \, q_0^4
\label{ERE} 
\end{equation}
with symmetric and real valued matrices $A$, $R$ and $P$. $A$ and $R$ 
correspond to the usual scattering length and effective range. 
We include here also the next term in the expansion, $P$, for testing purposes. 
By switching it on and off we can examine the stability of the pole positions.
The quantity $q$ in Eq.~(\ref{ERE}) 
is a diagonal matrix with the on-shell momenta of the individual channels.
$q_0$ is the momentum relative to the threshold at which the expansion
is performed. Note that the ERE of $M$ in Eq. (\ref{ERE}) 
can be also written in symmetric (matrix) form \cite{Badalyan:1982} utilizing diagonal
momentum matrices of the form $(q^2 - q^2_{(th)})^{1/2}$, etc.\ . Here $q_{(th)}$ are the
momenta $q_i$ for each channel that correspond to the energy of the
threshold at which the effective range expansion is performed \cite{deSwart:1962}. 
The two ways of writing the expansion are equivalent \cite{Badalyan:1982}.

\begin{figure*}
\begin{center}
\includegraphics[height=72mm]{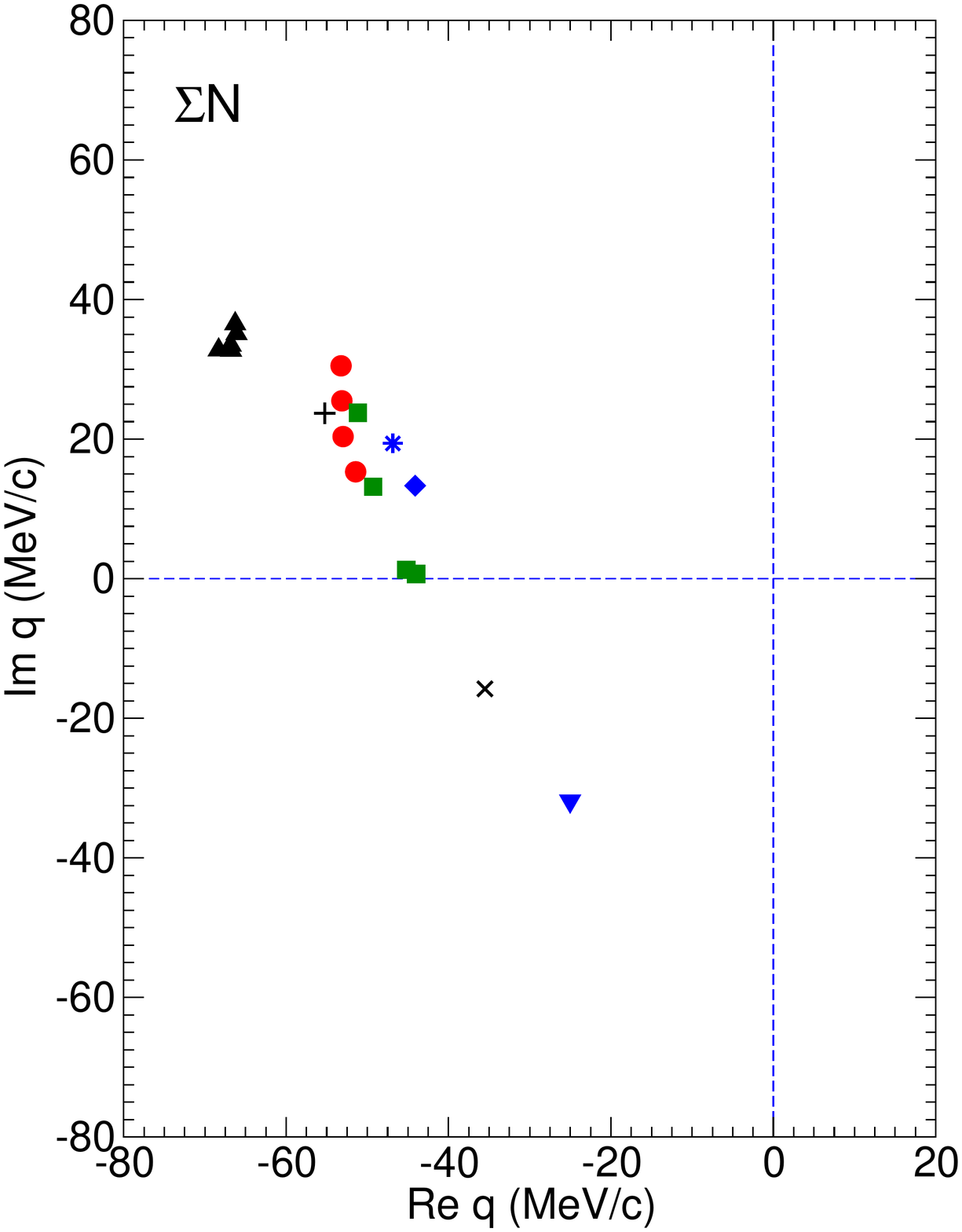}\includegraphics[height=72mm]{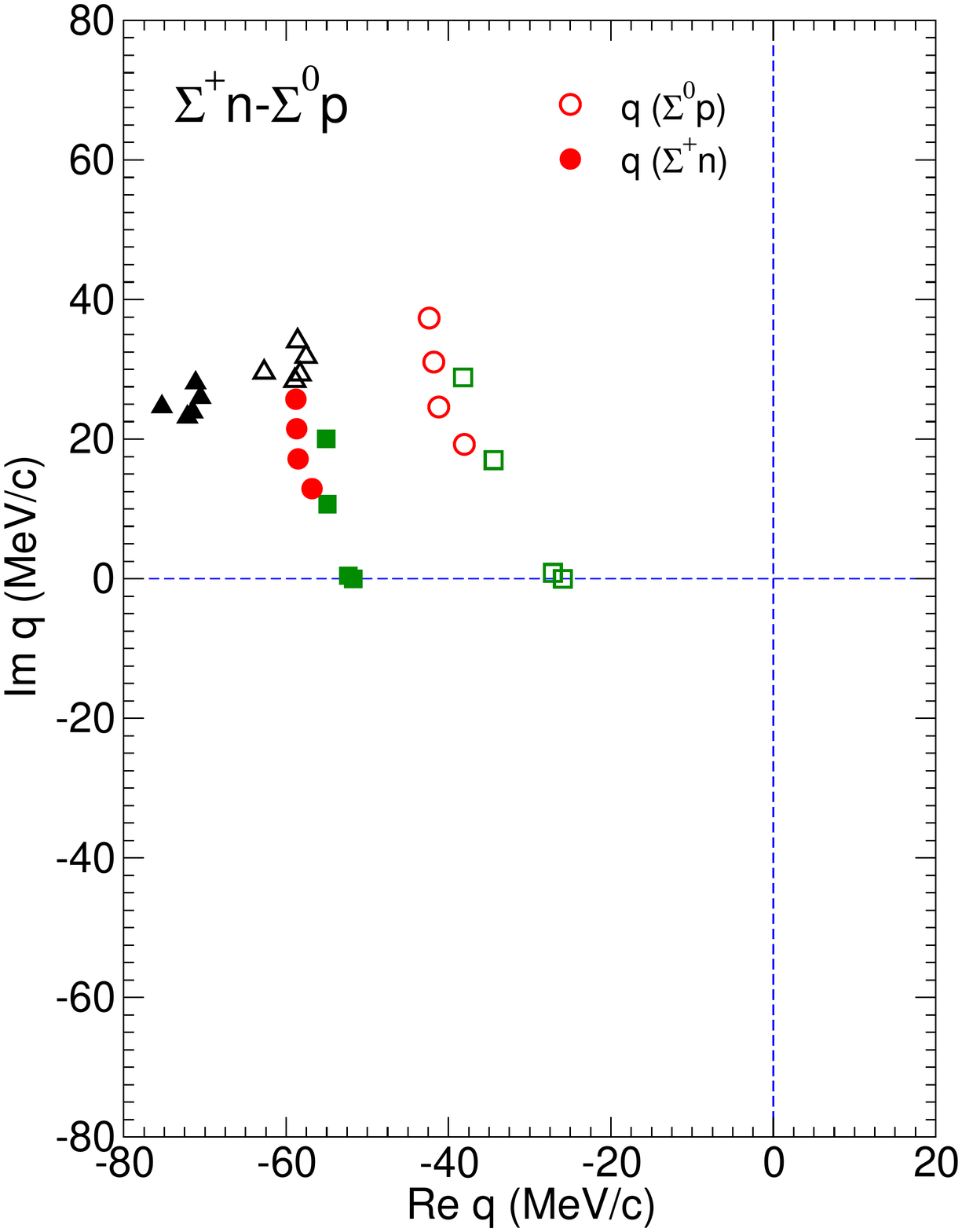}\includegraphics[height=72mm]{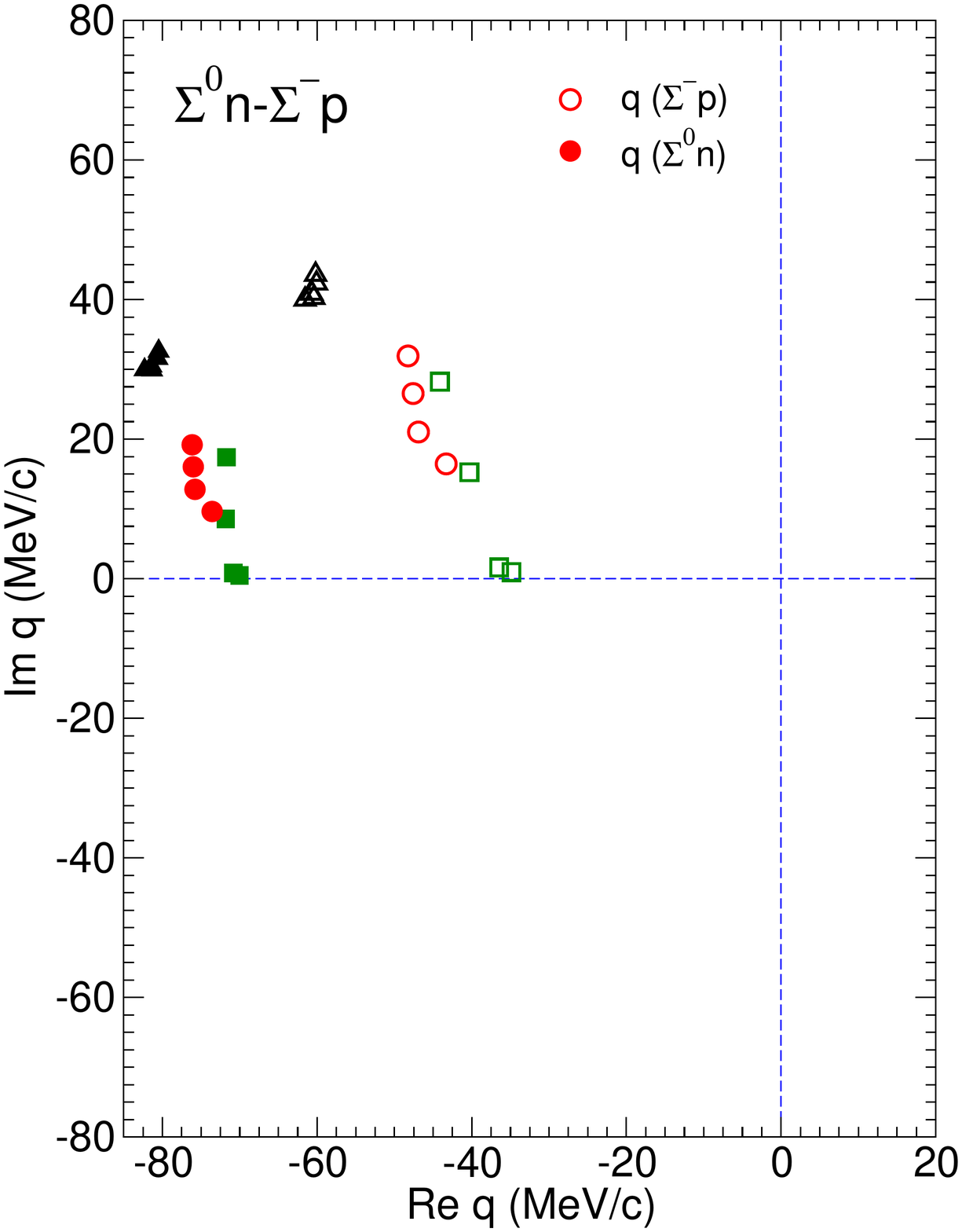}
\caption{$\Si N$ poles for isospin averaged masses (left), for $Q=1$ ($\Si^+ n$, $\Si^0 p$)
(middle) and for $Q=0$ ($\Si^0 n$, $\Si^- p$) (right). Results are shown for NLO19 (green squares),
NLO13 (red circles) and for  the Nijmegen NSC97b-f (black triangles) potentials. Filled (open)
symbols are for the $\Si N$ channel with the lower (higher) threshold. 
In case of isospin averaged masses also results for the J\"ulich meson-exchange 
potentials '04 \cite{Haidenbauer:2005} (blue inverted triangle), 
$\tilde A$ \cite{Reuber:1993} (blue diamond), and $A$ \cite{Holzenkamp:1989} (blue star), 
and the Nijmegen potentials ND \cite{Nagels:1977} (cross) and NF \cite{Nagels:1979} (plus)
are included. 
}
\label{fig:Poles}
\end{center}
\end{figure*}

Because of the tensor coupling of the $^3S_1$ and $^3D_1$ state, $F$ and accordingly
$M$ are $6\times 6$ matrices (or $4\times 4$ when isospin is conserved and 
the isospin $I=1/2$ system is considered). 
Near the $\Si N$ threshold the $^3D_1$ components of 
the two $\Si N$ channels are very small and can be safely neglected.
However, this is not the case for the $^3D_1$ $\La N$ partial waves which yields
an essential contribution at energies around the $\Si N$ threshold. Thus, 
our multichannel ERE involves $4\times 4$ matrices ($3\times 3$ in the isospin
symmetric case). 

An extensive discussion of the $\La N$-$\Si N$ coupled channels in terms of the effective
range approximation, within different scenarios, can be found in \cite{Badalyan:1982}. 
The expansion facilitates a reliable determination of the pole positions when they are 
within the analyticity circle bounded by the nearest dynamical singularity which 
for $\Si N$ is due to the left-hand cut caused by one-pion exchange. That cut
starts at $E = -m^2_{\pi}/8 \mu_{\Si N} \approx -5.1$~MeV 
below the $\Si N$ threshold \cite{Badalyan:1982}, so that $|q_0| \lesssim 70$~MeV/c.
In our study the expansion is performed at the higher of the two $\Si N$ thresholds, i.e.
at the one of the $\Si^0 p$ channel for $Q=1$ and of the $\Si^- p$ channel in case 
of $Q=0$. 
 
\begin{table*}
\renewcommand{\arraystretch}{1.5}
\centering
\begin{tabular}{|c|c|c|c|}
\hline
{NLO13} & {NLO19} & {J\"ulich '04} & NSC97b,d,f \\
\hline
\multicolumn{4}{|c|}{$Q=1$} \\   
\hline
$2131.90 -{\rm i\,}1.39$&  $2131.73 -{\rm i\,}1.11$&  $2129.01 +{\rm i\,}0.84$&  $2133.04 -{\rm i\,}3.80$ \\
$2131.92 -{\rm i\,}1.93$&  $2131.48 -{\rm i\,}2.10$&                         &   $2133.29 -{\rm i\,}3.25$ \\
$2131.62 -{\rm i\,}2.47$&  $2131.57 -{\rm i\,}0.04$&                         &   $2133.79 -{\rm i\,}3.53$ \\
$2131.25 -{\rm i\,}3.01$&  $2131.51 +{\rm i\,}0.00$&                         &                            \\
\hline
\multicolumn{4}{|c|}{$Q=0$} \\   
\hline
$2137.20 -{\rm i\,}1.35$&  $2136.99 -{\rm i\,}1.16$&  $2134.17 +{\rm i\,}0.57$&  $2137.31 -{\rm i\,}4.99$ \\
$2137.34 -{\rm i\,}1.87$&  $2136.77 -{\rm i\,}2.35$&                         &   $2137.58 -{\rm i\,}4.71$ \\
$2137.16 -{\rm i\,}2.40$&  $2136.93 -{\rm i\,}0.11$&                         &   $2137.75 -{\rm i\,}4.68$ \\
$2136.92 -{\rm i\,}2.93$&  $2136.82 -{\rm i\,}0.06$&                         &                            \\
\hline
\hline
\end{tabular}
\caption{Poles in the energy plane (in MeV). Listed are results for the NLO13 
\cite{Haidenbauer:2013} and NLO19 \cite{Haidenbauer:2019} potentials (for cutoffs 
$650$ to $500$~MeV from top to bottom)
and for the J\"ulich '04 \cite{Haidenbauer:2005} and Nijmegen NSC97 \cite{Rijken:1999} 
meson exchange potentials. The threshold for $Q=1$ are
$2128.97$ ($2130.87$)~MeV for $\Si^+ n$ ($\Si^0 p$), 
those for $Q=0$ are $2132.17$ ($2135.67$)~MeV for $\Si^0 n$ ($\Si^- p$).
} 
\label{tab:Epole} 
\end{table*}

Our results are summarized in Fig.~\ref{fig:Poles} where we show the $\Si N$ poles in 
the complex $q_{\Si N}$ plane for calculations with isospin-averaged masses (left),
for $Q=1$ (middle) and for $Q=0$ (right). 
In the presentation and discussion of the pole positions we adopt the definitions and conventions 
of Badalyan et al.~\cite{Badalyan:1982}. Specifically, we use the following classification
of sheets in the complex $q$ plane, 
\begin{eqnarray}
\nonumber
{\rm sheet\phantom{xx} I}&:& {\rm Im} \ q_{\La N} > 0, {\rm Im} \ q_{\Si N} > 0, \\
\nonumber
{\rm sheet\phantom{x,} II}&:& {\rm Im} \ q_{\La N} < 0, {\rm Im} \ q_{\Si N} > 0, \\
\nonumber
{\rm sheet\phantom{x} III}&:& {\rm Im} \ q_{\La N} < 0, {\rm Im} \ q_{\Si N} < 0, \\
\nonumber
{\rm sheet\phantom{x} IV}&:& {\rm Im} \ q_{\La N} > 0, {\rm Im} \ q_{\Si N} < 0 , 
\end{eqnarray}
appropriate for two channels. 
We focus here on the poles for the $\Si N$ channels. For all considered $YN$ interactions 
the poles are located either on sheet~II or IV, that correspond to the sheets [bt] and [tb] 
in another popular labeling scheme \cite{Pearce:1989}.
Specifically, they lie in the second or third quadrant in the complex $q_{\Si N}$ plane. 
Thus, in practice, they classify as unstable bound states (UBS) or 
inelastic virtual states (IVS) in the terminology of Ref.~\cite{Badalyan:1982}.
In principle, they correspond to what is termed coupled-channel (CC) poles in 
that work because their position is significantly influenced by the strong channel coupling
between $\La N$ and $\Si N$ and there are no poles near the $\Si N$ thresholds for the $YN$
interactions discussed here when the coupling is switched off. 
Accordingly, the essential question is whether the poles
appear and remain on sheet~IV in the full coupled-channel calculation or whether 
the $\Si N$ interaction together with the $\La N$-$\Si N$ coupling is strong enough 
so that the poles are on sheet~II.
 
Note that eight sheets appear in the three-channel case \cite{Badalyan:1982} and the 
notation has to be generalized accordingly. However, since in our study the poles 
for $\Si^+ n$ and $\Si^0 p$ ($\Si^0 n$ and $\Si^- p$) turn out to be always on the 
same respective sheets we refrain from introducing a more complicated notation 
and we show the poles in the same panel in Fig.~\ref{fig:Poles}. 

Miyagawa and Yamamura have determined the poles for some of the Nijmegen potentials 
directly by solving the scattering equation, analytically 
continued into the complex plane \cite{Miyagawa:1999}. Specifically, results for the 
NSC97f potential are provided. 
Our value of $q_{\Si N} = (-68.3,\,32.8)$~MeV/c for NSC97f compares rather well with 
$q_{\Si N} = (-69,\,30)$~MeV/c quoted in Table~1 of~\cite{Miyagawa:1999}, 
which gives us confidence that the ERE is quite reliable for establishing the position 
of the near-threshold poles. The agreement is particularly remarkable in view of 
the fact that the pole lies already close to the formal boundary where the ERE 
is expected to work reliably, see above. 

We start with the poles found for isospin-averaged masses, summarized in the left panel 
of Fig.~\ref{fig:Poles}. In this case isospin symmetry is fulfilled and the poles 
reflect directly the strength of the coupled-channel $\La N$-$\Si N$ interaction with $I=1/2$.
Evidently for all the NLO EFT interactions and also for the NSC97 potentials 
the poles lie on sheet~II, i.e. all of them predict a UBS \cite{Badalyan:1982}. 
It is certainly remarkable that the poles lie all in a narrow region. Thus,
despite of the large experimental uncertainties for some of the $YN$ cross sections,
the $\Si^- p$ data as a whole seem to impose rather strong restrictions. 
Indeed the potentials with the lowest $\chi^2$ (cf. Table~\ref{tab:chi1}) yield
also similar pole positions. 
Two of the NLO19 potentials, the ones with cutoff $500$ and $550$~MeV, stick out
because they yield poles where Im~$q_{\Si N}$ is very close to zero, i.e. the poles
are close to sheet~IV where the IVS are located. But in these two cases the
achieved $\chi^2$ is already noticeable larger as seen in Table~\ref{tab:chi1}.   
For illustration we include here the pole positions for other $YN$ interactions
like the J\"ulich '04 \cite{Haidenbauer:2005}, {$\tilde A$} \cite{Reuber:1993}, 
and $A$ \cite{Holzenkamp:1989} potentials. In addition, results for the 
Nijmegen potentials ND \cite{Nagels:1977} and NF \cite{Nagels:1979} 
(taken from Ref.~\cite{Miyagawa:1999}), are included.  
Among those only ND and J\"ulich '04 predict an IVS. 
But in the former case the $\chi^2$ is noticeable larger \cite{Nagels:1977}
than the best values in Table~\ref{tab:chi1}
because the $\Si^- p$ cross section is somewhat low 
and the latter yields a too low $\Si^- p\to \La n$ transition cross section
\cite{Haidenbauer:2013} and a too large capture ratio. 

When we use physical masses and consider the $Q=0$ and $Q=1$ systems 
the overall picture does not change qualitatively. 
This is not too surprising because the $\Si N$ interaction in the $^3S_1$-$^3D_1$
partial wave is dominated by the $I=1/2$ component. The $I=3/2$ contribution is 
small \cite{Rijken:1999} and, in case of the EFT 
potentials, the corresponding interaction is even weakly repulsive  
\cite{Haidenbauer:2013,Haidenbauer:2019}.
With the mass splitting taken into account, there are poles for 
$\Si^+ n$ and $\Si^0 p$, and for $\Si^0 n$ and $\Si^- p$, respectively, but
they all lie in the same quadrant as before, i.e. correspond again to a UBS.
The poles for the $\Si^0 n$ and $\Si^- p$ channels, displayed in 
the right-hand panel of Fig.~\ref{fig:Poles}, are somewhat closer together
than those for $\Si^+ n$ and $\Si^0 p$. It is a consequence of the fact 
that the $\Si N$ interaction is primarily determined by the scattering 
data in the $\Si^- p$ channel.
The results for Nijmegen NSC97 potentials are particularly close together.
Presumably, due to the absence of constraints on the relative strength 
of the singlet- and triplet $\La N$ interactions, an optimal description 
of the $\Si N$ data could be achieved for all versions a-f. 
Overall, the largest variations occur for NLO19, where, 
however, as said, here is also a larger difference in the achieved 
$\chi^2$, cf. Table~\ref{tab:chi1},

\begin{figure*}
\begin{center}
\includegraphics[height=65mm]{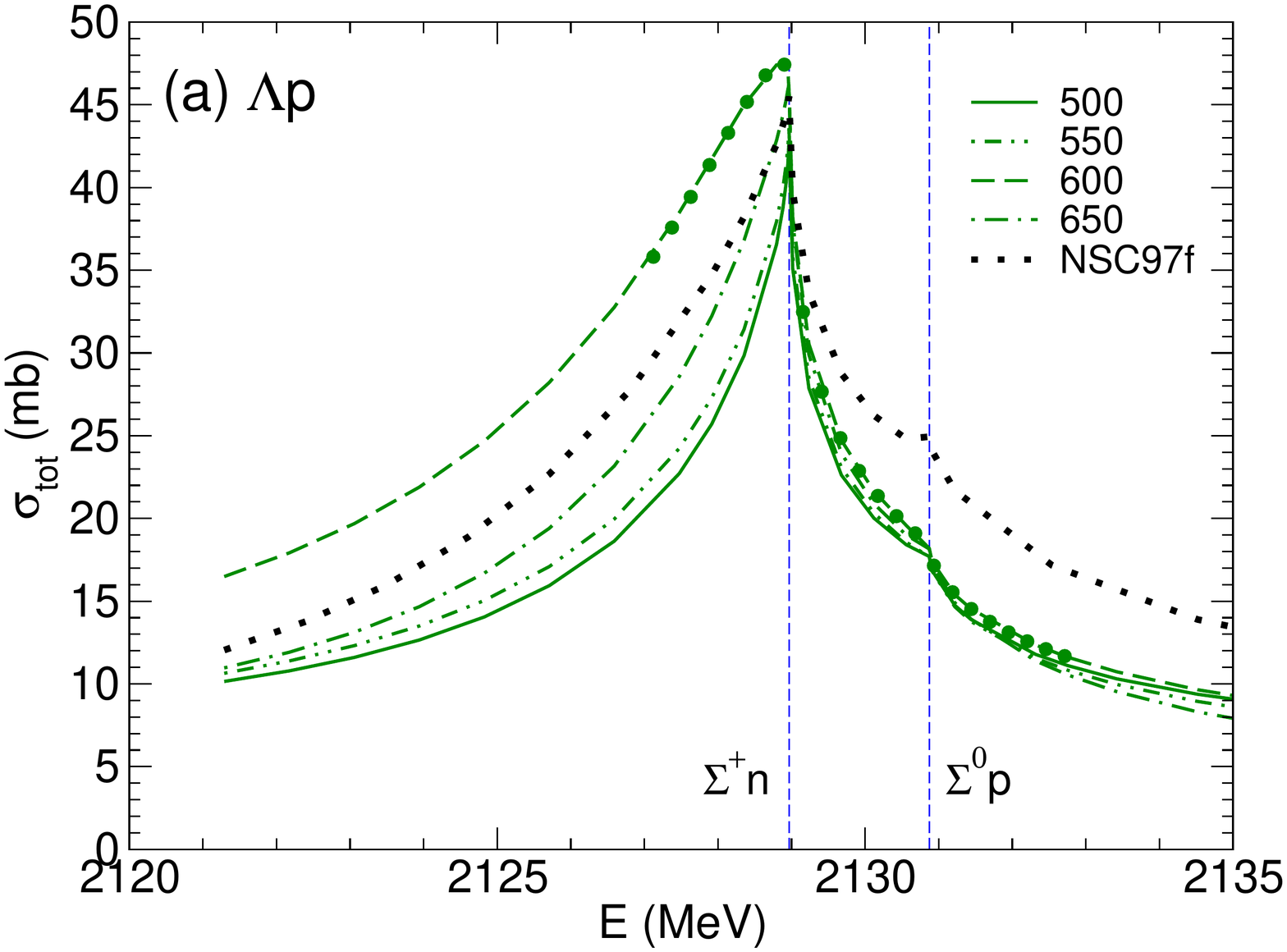}\includegraphics[height=65mm]{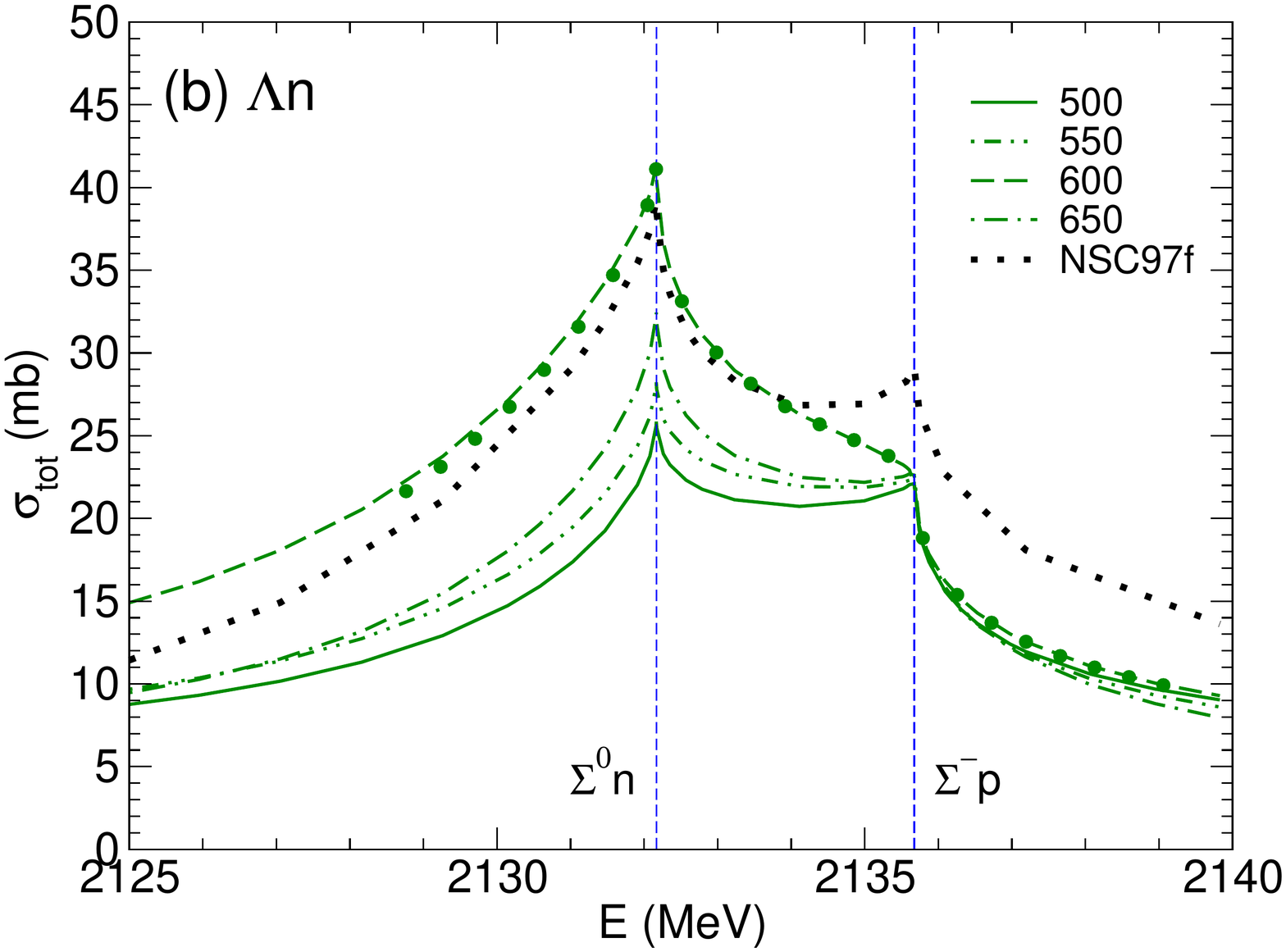}
\caption{Results for $\La p$ (a) and $\La n$ (b) cross sections for the NLO19 interaction with
cutoffs $500$-$650$~MeV and for the NSC97f potential. Circles indicate the 
results for NLO19~(600) based on the effective range expansion (\ref{ERE}). 
The vertical lines indicate the $\Si N$ thresholds. 
}
\label{fig:TT2}
\end{center}
\end{figure*}

As manifested by Fig.~\ref{fig:Poles}, for the EFT interactions and the
NSC97 potentials all poles lie in the lower half of the second quadrant of 
the complex $q_{\Si N}$ plane, so that Im~$q_{\Si N}$ $<$ -Re~$q_{\Si N}$. 
This means that the real part of the corresponding 
energy, $E=\sqrt{M^2_N+ q^2_{\Si N}} + \sqrt{M^2_{\Si}+ q^2_{\Si N}}$, 
is larger than $M_N+M_{\Si}$, i.e. the ``bound'' states lie actually above the 
$\Si N$ threshold. 
We summarize the pole positions in the energy plane in Table~\ref{tab:Epole}  
for the calculation with physical masses. 
As expected in all cases (except for the J\"ulich~'04 potential) 
the poles are indeed located above the $\Si N$ thresholds. 
 
In any case, we interpret our results as strong evidence 
for the existence of a dibaryon, in form of a (unstable) $\Si N$ bound state in the 
vicinity of the $\Si N$ threshold. It is supported by all $YN$ interactions that 
provide the best possible reproduction of the near-threshold $\Si N$ data.
In this context we note that the important role of the $\Si N$ data for the appearance 
of a $\La p$ resonance was already pointed out long time ago \cite{Fast:1969}.
Admittedly, what we obtain here is not the kind of dibaryon one ideally wants to
have. The position of the pole in the energy plane is above the $\Si N$ threshold. 
There is no Breit-Wigner type peak that is well separated from and well below 
the $\Si N$ thresholds. 
Furthermore, there is no ``pre-existing'' $\Si N$ bound 
state, i.e. there is no bound state when the $\La N$-$\Si N$ coupling is switched off. 
Nonetheless the position of the poles is in all cases in the second
quadrant of the complex $q_{\Si N}$ plane, i.e. where unstable bound states
are to be found \cite{Badalyan:1982}. 

\subsection{Shape of the $\La p$ and $\La n$ cross sections}
Detailed discussions of the shape of the $\La N$ cross section depending on 
the position of the poles can be found in 
Refs.~\cite{Pearce:1989,Miyagawa:1999,Afnan:1993}.
Here we focus on the shape of the $\La p$ and $\La n$ cross sections for the $YN$ 
potentials NLO19 and NSC97f, see Fig.~\ref{fig:TT2}. Indeed, the results for NLO13 
are well within the variations found for NLO19. Furthermore, the differences between 
the NSC97a-f predictions are rather small, so that concentrating on results for NSC97f  
is sufficient.
Generally speaking, a pole on sheet~IV (third quadrant of $q_{\Si N}$) leads to a 
cusp at the $\La N$ threshold while a pole on sheet~II (second quadrant of $q_{\Si N}$) 
produces a rounded (Breit-Wigner type) peak. In practice, this happens only for
ideal cases where the pole is sufficiently close to the $\Si N$ threshold and 
specifically close to the negative or positive Im~$q_{\Si N}$ axis. As one can see 
in Fig.~\ref{fig:Poles} the second condition is not fulfilled by any of the $YN$
potentials considered here. Thus, we are far from the mentioned ideal cases
and that means, citing Pearce and Gibson \cite{Pearce:1989}, in ``a gray area 
where it is not obvious whether the effect will be cusp or peak''. 

Starting with the $\La n$ cross section at the $\Si^0 n$ threshold 
(Fig.~\ref{fig:TT2}b) one
can see that a cusp is produced by all potentials, despite of the fact that 
the poles for NSC97f and for the NLO19 potentials lie on sheet~II. In case of 
NLO19~(550) and NLO19~(500) the poles
are very close to the Re~$q_{\Si N}$ axis which separates sheet~II and IV, 
so that a cusp is may be not too surprising. At the $\Si^- p$ threshold
some potentials produce a cusp while others lead to a rounded step. The
latter is the expected behavior based on the pole position. In any case,
the presence of a lower near-by threshold distorts the signal strongly 
so that no rounded peak appears. 

In case of the $\La p$ cross section (Fig.~\ref{fig:TT2}a) most of the potentials produce again
a cusp at the lower $\Si N$ threshold, despite of having the poles on sheet~II. 
The only exception is NLO19~(600) where a rounded peak is visible, at least 
on the scale chosen for the figure. Note that the peak is barely $100$~keV below
the threshold and certainly is not of Breit-Wigner type. A similar shape
was reported in Ref.~\cite{Miyagawa:1999} for the NSC97f potential in a
calculation using isospin-averaged masses. 
At the $\Si^0 p$ threshold cusps as well as rounded steps occur. However, 
the actual signals are obscured by the large effect at the $\Si^+ n$ 
threshold and the subsequent steep fall-off of the $\La p$ cross section.

\begin{table*}
\caption{Hadronic shifts and broadenings of $S$-wave states of $\Si^-$p atoms (in eV).
Results for $^1S_0$ and $^3S_1$ partial waves and for the spin average are given.
}
\renewcommand{\arraystretch}{1.2}
\label{tab:levelC}
\vspace{0.2cm}
\centering
\begin{tabular}{|c|rrrr|rrrr|r|r|}
\hline
& \multicolumn{4}{|c|}{NLO13} & \multicolumn{4}{|c|}{NLO19} & J\"ulich '04 & NSC97f  \\
\hline
{$\Lambda$ (MeV)} & $500$ & $550$& $600$& $650$& 
 $500$ & $550$& $600$& $650$  & & \\
\hline
\hline
$E_{^1S_0}$       &$-$248  &-231  &-146  & $-$106      & $-$249 & -234 & -146  & $-$107   & $-$130  & $-$498  \\
$\Gamma_{^1S_0}$  &1401    &1391  & 1357 &   1317      & 1471   & 1455 & 1381  &  1309      &  1788   &  1809   \\
$E_{^3S_1}$       &$-$1286 &-1256 &-1211 & $-$1159     & $-$944 & -942 & -1210 & $-$1141 & 884     & $-$825  \\
$\Gamma_{^3S_1}$  & 2338   &2514  &2657  & 2865        & 3506   & 3406 & 2620  & 2975       &  4782   &  2605  \\
$E_{1S}$          &$-$1026 &-1000 &-945  & $-$896      & $-$770 & -765 & -944  & $-$882   &  630    & $-$743  \\
$\Gamma_{1S}$     &2104    & 2233 &2332  & 2478        & 2997   & 2918 & 2310  &   2558     &  4034   &  2406  \\ 
\hline
\hline
\end{tabular}
\renewcommand{\arraystretch}{1.0}
\end{table*}

Since for $\La n$ the separation of the $\Si N$ thresholds 
is significantly larger than for the $Q=1$ channels, the details
of both threshold structures appear more prominently. 
Moreover, the interaction in the $\Si^- p$ channel is stronger
than that for $\Si^0 n$ because the crucial $I=1/2$ contribution
enters with a weight $2/3$ in the former and with $1/3$ in the
latter \cite{Holzenkamp:1989}. For $Q=0$ the contribution to the lower 
channel, $\Si^+ n$, is weighted by $2/3$.
Both aspects make the $\La n$ channel to be a good testing ground for
details of the $YN$ interaction. Unfortunately, there is little 
hope to perform pertinent experiments.
In any case, we want to emphasize that in the $\Si^- p$ channel 
the real situation will be more complicated,
because the presence of the attractive Coulomb interaction
leads to an accumulation of Coulomb bound states at the
threshold. In this case there is a discontinuity of the cross sections at the 
$\Si^- p$ threshold and, as a consequence, no cusp (or rounded step) is
expected but a jump in the cross sections of the open channels. For a
detailed discussion see Ref.~\cite{Newton:1966}.
We do not consider this complication here. 

Finally, in order to demonstrate the quality of the ERE, we indicate in 
the figure corresponding results for NLO19~(600) (circles). As one can see, 
the representation of the amplitudes in terms of such an expansion works 
remarkably well, down to and even below the lower $\Si N$ threshold. 

\section{Level shifts and widths of $\Si^- p$ atoms}

Measurements of $\Si^- p$ scattering with reduced uncertainty would be 
rather useful for corroborating the existence of a $S=-1$ dibaryon 
suggested by the present study. 
An alternative source of information is offered by
measurements of level shifts and widths of $\Si^- p$ atoms. 
In Table~\ref{tab:levelC} we present predictions for those
quantities, for the $YN$ potentials considered in the present work.  
In the calculation the Trueman formula \cite{Trueman:1961} was
applied (with the second-order term taken into account) 
which relates these quantities to the $\Si^- p$ scattering lengths: 
\begin{equation}
\Delta E + {\rm i}\, {\frac{\Gamma}{2}} = - \frac{2}{\mu_{\Si^- p} r^3_B} {a^{sc}} 
\left(1-\frac{a^{sc}}{r_B} \beta\right) 
\end{equation}
Here ${a^{sc}}$ is the Coulomb-modified $\Si^- p$ ($^1S_0$ and/or $^3S_1$) 
scattering length, $\mu_{\Si^- p}$ is the reduced mass, and $r_B$ is the Bohr 
radius which amounts to $51.4$~fm for $\Si^- p$. 
The quantity $\beta$ is given by $\beta = 2(1-\Psi(1)) \approx 3.1544$ 
for $S$ waves, where $\Psi$ is the digamma function. 
According to a detailed study of antiprotonic atoms \cite{Carbonell:1992} the above
formula yields rather reliable results once the Coulomb interaction is explicitly
taken into account in the calculation of the hadronic reaction amplitude and/or
scattering length.
 
It is interesting to compare our results with those for similar 
atomic systems where measurements have been already performed. 
This is possible for antiprotonic hydrogen and deuterium \cite{Gotta:2004} 
as well for kaonic hydrogen \cite{Bazzi:2011}. It reveals that the predicted 
widths for $\Si^- p$, being of the order of ($2100$-$3000$)~eV, are noticeable 
larger than those for antiprotonic 
atoms ($\Gamma_{\bar pp} \approx 1000$~eV) \cite{Gotta:2004}  
and $K^-$ atoms ($\Gamma_{K^- p} \approx 500$~eV) \cite{Bazzi:2011}. 
Most likely this is due to the fact that the threshold of the neutral 
``partner'' channel ($\Si^0 n$) is slightly below the one of $\Si^- p$
whereas in the other two systems the corresponding channels 
($\bar nn$ and $\bar K^0 n$) are slightly above. 
In any case a large width as predicted here certainly reduces the prospects
for an experimental determination of the level shifts and widths of $\Si^- p$ atoms. 
So far such measurements have been only performed for carbon 
and heavier nuclei \cite{Batty:1978,Batty:1994}. 

\section{Summary}

In the present work we studied the threshold structure seen in the $\La p$ 
cross section (invariant mass spectrum) around the $\Si N$ threshold. For
that purpose we utilized $YN$ interactions that yield the presently 
best description of low-energy $\La p$, $\Si^- p$ and $\Si^+ p$ scattering data.
The $YN$ potentials in question are interactions established within chiral effective 
field theory up to next-to-leading order by the J\"ulich-Bonn-Munich
group in 2013 and 2019 \cite{Haidenbauer:2013,Haidenbauer:2019} and 
the Nijmegen NSC97 meson-exchange potentials \cite{Rijken:1999} from 1999. 
In all those cases the achieved $\chi^2$ value is in the order of $16$ for 
the $36$ (or $35$) ``best'' $YN$ data taken into account.

Our work revealed that 
(i) if one takes into account the full complexity of the $YN$ interaction 
(tensor forces, $\La N$-$\Si N$ coupling) as well as constraints from 
(broken) SU(3) flavor symmetry \cite{Haidenbauer:2013,Haidenbauer:2019,Rijken:1999} 
and  
(ii) one takes the presently available low-energy $\Si N$ data serious 
and aims at their best possible reproduction, 
then the appearance of a dibaryon in form of a deuteron-like (unstable) $\Si N$ 
bound state seems to be practically unavoidable. 

Unfortunately, our study also indicates that it might be wishful thinking to 
expect a truly convincing evidence for a strangeness $S=-1$ dibaryon, i.e. 
a peak that is well separated from the (and well below the) $\Si N$ threshold.  
Nonetheless, to confirm our result and to reliably establish that there
is a pole in the second quadrant of the complex $\Si N$ momentum plane
which signals a $\Si N$  bound state,
additional and more accurate near-threshold $\Si^- p$ data would
be rather useful. 
It would be also interesting to get experimental data or at least tighter
constraints on the charge $Q=1$ $\Si N$ channels. Pertinent information 
has been already acquired at the COSY accelerator in J\"ulich, 
from the reactions
$pp \to K^+ \Si^0 p$ \cite{Sewerin:1998,Kowina:2004}
and $pp \to K^+ \Si^+ n$ \cite{Rozek:2006,Valdau:2007,Valdau:2010},
and also by the ALICE Collaboration where the $\Si^0 p$ momentum
correlation function was determined in $pp$ collisions at $13$~TeV
\cite{Acharya:2020}. But the present quality of the data
together with uncertainties in the tools for analyzing final-state
interactions \cite{Gasparyan:2003,Haidenbauer:2018} prevent more 
quantitative conclusions. 
Most promising are certainly planned scattering experiments at J-PARC,
where among other things the reactions $\La p \to \Si^0 p, \Si^+ n$ could
be measured \cite{Miwa}. Such cross sections would provide 
independent information on the $\La N\leftrightarrow\Si N$
transition, complementing available data for $\Si^- p \to \La n$,
and, thus, could allow one to pin down the actual strength of the
$\La N$-$\Si N$ coupling more accurately. 

\section*{Acknowledgements}
We would like to thank Michael D\"oring for 
the code for searching for zeros in the complex plane.
This work is supported in part by  the 
Deutsche Forschungsgemeinschaft (DFG, German Research Foundation)
and the NSFC through the funds provided to the Sino-German Collaborative
Research Center  CRC~110 ``Symmetries and the Emergence of Structure in QCD''
(DFG Project-ID 196253076 - TRR 110, NSFC Grant No. 12070131001)
and by the EU (STRONG2020).
The work of UGM was supported in part by the Chinese
Academy of Sciences (CAS) President's International
Fellowship Initiative (PIFI) (Grant No. 2018DM0034)
and by VolkswagenStiftung (Grant No. 93562).

\end{document}